\documentclass[aps,prl,twocolumn]{revtex4-1}

\usepackage{amsmath}
\usepackage{amsfonts}
\usepackage{amssymb}
\usepackage{color}

\usepackage{graphicx}
\usepackage{xcolor}

\usepackage{bbold}
\usepackage{dsfont}

\newcommand{\be}{\begin{equation}}
\newcommand{\ee}{\end{equation}}

\newcommand{\bea}{\begin{eqnarray}}
\newcommand{\eea}{\end{eqnarray}}

\renewcommand{\Im}{{\rm Im}\,}
\renewcommand{\Re}{{\rm Re}\,}

\renewcommand{\epsilon}{\varepsilon}

\begin{document}
\date{\today}
\author{Ervand Kandelaki and Mark S. Rudner}
\affiliation{Niels Bohr International Academy and Center for Quantum Devices, Copenhagen University, Blegdamsvej 17, 2100 Copenhagen, Denmark}
\title{Many-body dynamics and gap opening in interacting periodically driven systems}
\begin{abstract}
We study the transient dynamics in a two-dimensional system of interacting Dirac fermions subject to a quenched drive with circularly polarized light.
In the absence of interactions, the drive opens a gap at the Dirac point in the quasienergy spectrum, inducing nontrivial band topology.
Here we investigate the dynamics of this gap opening process in the presence of interactions, as captured by the generalized spectral function and correlators probed by photoemission experiments.
Through a mechanism akin to that known for equilibrium systems, interactions renormalize
and enhance the induced gap over its value for the non-interacting system.
We additionally study the heating that naturally accompanies driving in the interacting system, and discuss the regimes where dynamical gap emergence and enhancement can be probed before heating becomes significant.
\end{abstract}
\maketitle
{\it Introduction.---}
Over the past decade, periodic driving has attracted intense interest as a means for controlling and investigating a variety of intriguing quantum many-body phenomena~(see, e.g., Refs.~\cite{Yao2007,Fistul2007,Syzranov2008,Oka2009,Kitagawa2010,Lindner2011,Kitagawa2011,Gu2011,Jiang2011,Kundu2013,Iadecola2013,Grushin2014,Genske2015,Quelle2016,Khemani2016,vonKeyserlingk2016, Potter2016,Else2016,Roy2016,Dutreix2016,DeGiovannini2016,Huebener2017,Stepanov2017,Else2017}).
In particular, driving has been proposed as a new route for altering the topological properties of Bloch bands, opening the possibility for dynamically switching between trivial and topological regimes~\cite{Oka2009,Kitagawa2010,Lindner2011}.
Along with considerable activity on the theoretical side,
remarkable progress has been achieved in the experimental realization of these systems in cold atomic, optical, and solid state systems~\cite{Aidelsburger2013,Jotzu2014,Lohse2016, Nakajima2016,Rechtsman2013,Hu2015,Wang2013,Mahmood2016,Choi2017,Zhang2017}.

In equilibrium, the topology of a band insulator's Bloch bands dictates the system's behavior at low temperatures.
Due to the system's energy gap, the presence of weak interactions does not qualitatively change the system's properties.
However, renormalization of parameters due to interactions may lead to potentially important {\it quantitative} changes, e.g., by enhancing the many-body gap above its noninteracting value~\cite{Kane2005,Kotov2008,Kotov2012,Song2013,Ugeda2014}.
The role of interactions in non-equilibrium driven systems is much more subtle.
While interactions may allow for interesting correlations to build up~\cite{Eckardt2005,Tsuji2008,Grushin2014,Mikami2016,Mitrano2016,Knap2016,Sentef2016,Klinovaja2016,Thakurathi2017,Coulthard2017,Qin2017}, they also provide pathways for the system to absorb energy from the driving field, and thereby to heat up~\cite{DAlessio2014,Lazarides2014,Bilitewski2015,Bukov2016}.
Thus understanding the interplay between these phenomena and the parameters that control them is crucial for enabling further advances in the field.

Recently, various properties of non-interacting Floquet systems have been explored extensively, such as their topological characteristics~\cite{Oka2009,Kitagawa2010,Kitagawa2011,Lindner2011,Gu2011,Jiang2011,Kundu2013, Rudner2013,Perez-Piskunow2014,Usaj2014,Nathan2015,vonKeyserlingk2016,Else2016, Potter2016,Roy2016},
and the relaxation dynamics associated with coupling to external reservoirs~\cite{Galitskii1970,Tsuji2009,Iadecola2013,Dehghani2014,Dehghani2015,Iadecola2015,Iadecola2015a, Seetharam2015,Liu2015,Shirai2015,Shirai2016,Iwahori2016,Liu2017}.
In strongly disordered, {\it closed} systems, many-body localization (MBL) provides a mechanism to avoid heating and to stabilize a variety of interesting phases~\cite{Basko2006,Gornyi2005,Ponte2015,Lazarides2015,Khemani2016,vonKeyserlingk2016,Else2016,Roy2016,Potter2016,Else2017,Bairey2017,Roy2017,Po2016,Bordia2017}.
Outside the MBL regime, periodically-driven systems also exhibit interesting {\it transient} dynamics~\cite{Wang2013,Mitrano2016,Knap2016,Sentef2016,Coulthard2017,Bukov2015,Eckardt2015,Else2017,Abanin2017,Lindner2017,Zeng2017}.

In the solid state context, the opening of
Floquet gaps on the surfaces of three-dimensional topological insulators via circularly polarized light has been investigated experimentally via transient pump-probe photoemission experiments~\cite{Wang2013,Mahmood2016}.
A detailed theory has been developed to describe the expected time-resolved Angular-Resolved Photoemission Spectroscopy (TRARPES) signal,
both for an isolated system~\cite{Sentef2015} and for a system coupled to a momentum-conserving phonon bath~\cite{Dehghani2014,Dehghani2015}, in the absence of electron-electron interactions.

\begin{figure}[t]
        \includegraphics[width=\columnwidth]{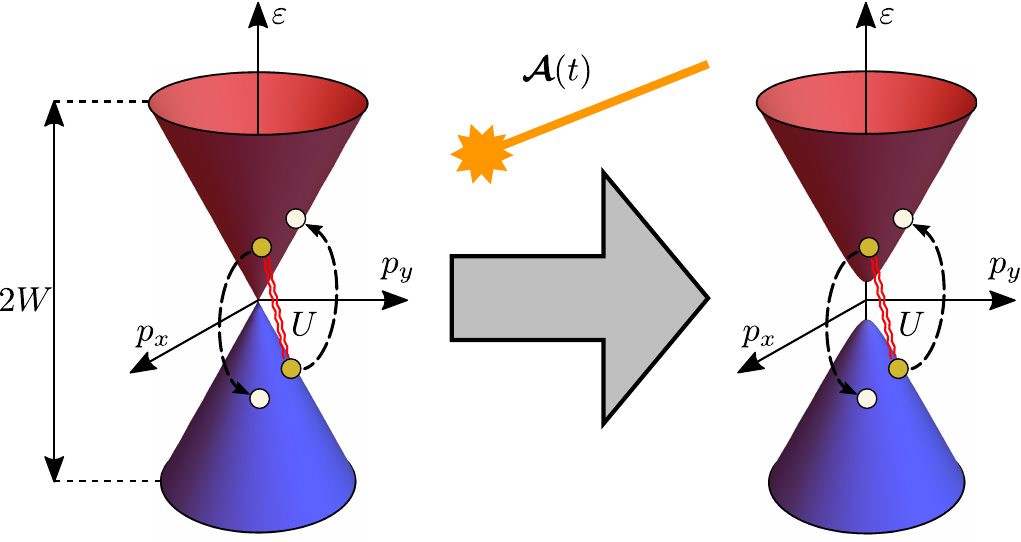}
\caption{We consider a single species of interacting 2D Dirac fermions, with finite bandwidth, $2\sqrt{2}W$.
A normally-incident circularly polarized driving field, described by the vector potential $\boldsymbol{\mathcal{A}}(t)$, opens a gap in the single-particle spectrum.
The strength of the local interaction is denoted by $U$.
}
\label{fig:model_cartoon}
\end{figure}

In this work we aim to elucidate the transient dynamics of driving-induced gap opening in interacting fermionic systems.
We take a model of a single two-dimensional Dirac mode with finite bandwidth, subjected to irradiation by normally-incident circularly polarized light, see Fig.~\ref{fig:model_cartoon}.
In the high frequency driving regime, where the drive frequency $\Omega$ exceeds the bandwidth of the fermionic system, the primary role of the driving field at the single particle level is to open a gap in the Dirac spectrum (with magnitude proportional to $\mathcal{A}^2/\Omega$, where $\mathcal{A}$ is the amplitude of the drive).
Here we anticipate three main effects of repulsive interactions:
1) based on the insights from the renormalization analysis in equilibrium~\cite{Kotov2012}, we expect that the driving-induced gap (suitably defined) will be renormalized upwards to a value larger than the single-particle gap defined above,
2) interactions will induce dephasing, providing an intrinsic timescale for the gap to emerge, and
3) the system will absorb energy, eventually heating toward an infinite temperature-like state.
For $\Omega/W \gg 1$ the energy absorption rate is expected to be exponentially suppressed, opening a long time-window during which ``pre-thermal'' dynamics can be observed~\cite{Abanin2015,Eckardt2015,Bukov2015,Kuwahara2016,Abanin2017}.
For lower frequencies, $\Omega/2W \lesssim 1$, the drive resonantly couples states of the two Dirac bands.
We will focus on the regime $\Omega/2W \gtrsim 1$. (Note that we set $c,\hbar = 1$ throughout.)

We describe the dynamics via the Keldysh many-body Green's functions formalism~\cite{Rammer2007,Stefanucci2013}.
In order to capture the gap renormalization as well as heating we employ a self-consistent approach, approximating the self-energy by diagrams up to second order in the interparticle interaction.
Using the Keldysh Green's functions we compute a generalized spectral function and the correlators accessible in TRARPES experiments; we use these observables to define a notion of a gap in the interacting, non-equilibrium system, and describe the emergence of this gap in time.
When interactions are strong, the gap can be considerably enhanced as compared to its non-interacting value.
Within the parameter regime studied ($\Omega/2W \gtrsim 1$), we find that the energy absorption rate increases with interaction strength $U$ as $\gamma \sim U^{2}$.

{\it Problem setup.---}
At the single particle level, we begin with the Hamiltonian of a two-dimensional (2D) massless Dirac mode subjected to a time-dependent optical field:
\begin{align}
  \!\!\!\!\!H_{0}(t) & = v \sum_{\boldsymbol{p}} \sum_{\alpha\alpha'} c_{\boldsymbol{p}\alpha}^{\dagger}\,[\boldsymbol{p} + e \boldsymbol{\mathcal{A}}(t)] \cdot \boldsymbol{\sigma}_{\alpha\alpha'}\,  c_{\boldsymbol{p}\alpha'}, \label{eq:hamiltonian_nonint}
\end{align}
where $c_{\boldsymbol{p}\alpha}^{\dagger}$ ($c_{\boldsymbol{p}\alpha}$) creates (annihilates) a fermion with momentum $\boldsymbol{p}$ and spin $\alpha = \{\uparrow,\downarrow\}$,
$v$ is the Fermi velocity, $-e$ is the electron charge, and $\boldsymbol{\sigma}=\{\sigma^{x},\sigma^{y}\}$ is the vector of Pauli matrices.
The AC driving is incorporated through the vector potential $\boldsymbol{\mathcal{A}}(t)$. We impose a finite single-particle bandwidth
via a momentum cutoff, $p_x, p_y \in [-\tfrac{W}{v},\tfrac{W}{v}]$, where $2\sqrt{2}W$ is the bandwidth.

In this work we consider a spin-independent, local, density-density interaction.
The full interacting Hamiltonian takes the form $H(t) = H_0(t) + H_{\rm int}$, with
\begin{align}
  \label{eq:Hint}H_{\mathrm{int}}
                            &
                    = \frac{U}{2N} \sum_{\boldsymbol{p},\boldsymbol{p}',\boldsymbol{q}} \sum_{\alpha,\alpha'}
                        c_{\boldsymbol{p}-\boldsymbol{q},\alpha}^{\dagger} c_{\boldsymbol{p}'+\boldsymbol{q},\alpha'}^{\dagger}
                        c_{\boldsymbol{p}',\alpha'} c_{\boldsymbol{p},\alpha},
\end{align}
where $N \propto L^2$
is the system size (number of allowed $\boldsymbol{p}$-points in a finite-sized system).

{\it Observables.---}
We now describe the 
many-body dynamics following a quench in which a circularly polarized driving field is suddenly turned on at time $t = t_{0}$:  $\boldsymbol{\mathcal{A}}(t) = \mathcal{A} \theta(t-t_{0}) \big( \cos(\Omega t),\, \sin(\Omega t) \big)$, where $\mathcal{A}$ is the driving amplitude and $\theta(t)$ is the Heaviside step function.
We characterize the ensuing dynamics in terms of the TRARPES signal~\cite{Sentef2015}, proportional to
\begin{align}
    \!\!\! I_{\boldsymbol{p}}(\omega,t)
    \!=\! \Im\Big\{
    & \!\int \!\!\mathrm{d}t_{1} \!\! \int \!\!\mathrm{d}t_{2} \, e^{i \omega \tau}
    s(t_1) s(t_2) \operatorname{tr}[ G^{<}_{\boldsymbol{p}}(t_{1},t_{2}) ] \Big\},\!\!\! \label{eq:trARPES_def}
\end{align}
which is related to the lesser Green's function $G^{<}$ defined in Eq.~(\ref{eq:Gls_def}) below.
Here $\tau=t_1-t_2$, and $s(t_i)$ is the probe pulse profile, with $i = \{1,2\}$; for simulations we take a normalized Gaussian, $s(t_i)=\exp\big(-\frac{(t_{i}-t)^2}{2\sigma^2}\big)/\sqrt{2\pi\sigma^2}$.

To help elucidate the nature of the non-equilibrium dynamics, we also compute the generalized spectral function that would be accessible through a hypothetical tunneling experiment~\footnote{While a tunneling experiment may not be feasible for systems driven at optical frequencies, the generalized spectral function is informative and may be accessible in other implementations such as in cold atom systems.}.
Akin to the equilibrium case, the differential conductance can be expressed as
$g(\omega,t)=\int\! \mathrm{d}\boldsymbol{p} \,\nu_{\boldsymbol{p}}(\omega,t)$, where the generalized spectral function
\begin{align}
  \!\!\!\!
  \nu_{\boldsymbol{p}}(\omega,t) = - 2 \, \Im \Big\{
  \!\! \int \!\! \mathrm{d}\tau e^{i \omega \tau}
  \operatorname{tr}[ G^{R}_{\boldsymbol{p}}(t,t-\tau) ] \Big\} \label{eq:gen_Aspec_def}
\end{align}
is related to the retarded non-equilibrium Green's function, $G^{R}$ [see Eq.~(\ref{eq:GR_from_GS})].
In Eqs.~(\ref{eq:trARPES_def}) and
(\ref{eq:gen_Aspec_def}), $\operatorname{tr}$ denotes the trace over spin indices.

Finally, we investigate the energy absorption rate
\begin{align}
  \!\!\gamma
    & = \frac{d \langle H(t) \rangle} {dt}
    = 2 A W \Omega  \sum_{\boldsymbol{p}} \Re \Big\{ e^{i \Omega t} \operatorname{tr}[ \sigma^{-} G^{<}_{\boldsymbol{p}}(t,t) ] \Big\}. \label{eq:gamma_def}
\end{align}
The dimensionless parameter $A = e v \mathcal{A}/W$ describes the strength of the drive.
Integrated over time, $\gamma(t)$ gives the total energy absorbed
due to the drive, $\delta E(t)$.

Using the index notation ${j}\equiv(\boldsymbol{p},\alpha)$, the lesser Green's function $G^{<}$ appearing in Eqs.~(\ref{eq:trARPES_def}) and (\ref{eq:gamma_def}) is defined as
\begin{gather}
  {G}^{<}_{{j},{j}'}(t,t') =        i\langle c_{{j}'}^{\dagger}(t') c_{{j}}(t) \rangle_{\rho_0}. \label{eq:Gls_def}
\end{gather}
Here $c^\dagger_{j'}(t')$ and $c_j(t)$ are the creation and annihilation operators in a Heisenberg picture based at time $t = 0$; i.e., $c_{{j}}^{(\dagger)}(t) = \mathcal{U}^{\dagger}(t) c_{{j}}^{(\dagger)} \mathcal{U}(t)$ with
$\mathcal{U}(t) = \mathcal{T} e^{- i\int_{0}^{t} \!\mathrm{d}t' H(t')}$, where $\mathcal{T}$ is the time ordering operator.
The operator averages $\langle X \rangle_{\rho_0} = \operatorname{Tr}[ X \rho_{0} ]$ are taken with respect to the many-body density matrix $\rho_0$ (the state at $t = 0$);  $\operatorname{Tr}$ stands for the trace in the Fock space.
The retarded Green's function ${G}^{R}$ in Eq.~(\ref{eq:gen_Aspec_def}) is given by
\begin{gather}
  {G}^{R}_{{j},{j}'}(t,t') =  \theta\big(t-t'\big) (G^{>} - G^{<})_{{j},{j}'}(t,t'), \label{eq:GR_from_GS}
\end{gather}
with ${G}^{>}_{{j},{j}'}(t,t') = - i\langle c_{{j}}(t) c_{{j}'}^{\dagger}(t') \rangle_{\rho_0}$.

The Green's functions satisfy the Dyson equation
\begin{align}
  \hat{G} = \hat{G}_{0} + \hat{G}_{0} \circ \hat{\Sigma} \circ \hat{G}.
  \label{eq:dyson_RAK}
\end{align}
For $X = \{G_0,G,\Sigma\}$ (i.e., for the bare and full Green's functions and the self-energy), we use a $2 \times 2$ matrix form
\begin{align}
    \hat{X} & = (\hat{X}_{\eta\eta'}) =
                        \begin{pmatrix}
                            {X}^{R}     & {X}^{K}\\
                            0           & {X}^{A}
                        \end{pmatrix},\label{eq:RAK_matrix_def}
\end{align}
with ${X}^{A}(t,t') = [X^{R}(t',t)]^{\dagger}$ (advanced part), and ${X}^{K} = X^{<} + X^{>}$ (Keldysh part). The generalized matrix product 
``$\circ$'' in Eq.~(\ref{eq:dyson_RAK}) extends over $\alpha$, $\boldsymbol{p}$, $\eta$ and $t$ indices~\cite{Rammer2007}.

In practice, Eq.~(\ref{eq:dyson_RAK}) cannot be solved exactly; the self-energy $\Sigma$ must be approximated in an appropriate way to capture the relevant physics in the system.
While the simplest (Hartree-Fock) level approximation captures gap renormalization, it misses the effects of correlations and dissipative processes that destabilize the Floquet system (heating).
We thus use the second-Born approximation (2BA), calculating the self-energy self-consistently as a functional of the full Green's function, including all skeleton diagrams up to second order in the bare interaction.
The first order diagrams would yield the Hartree-Fock (HF) approximation and the second order diagrams include (some) correlation effects.
This approach is a ``conserving approximation,'' and thus our results respect macroscopic conservation laws~\cite{Rammer2007,Stefanucci2013}.

Within the 2BA, the only contribution to $\Sigma^{\gtrless}$ is of second order in the interaction strength, $U$:
\begin{align}
    \Sigma^{(2)\gtrless}_{\boldsymbol{p}} (t,t')
        & =
        \tfrac{U^{2}}{N}
        \sum_{\boldsymbol{p}'}
      \bar{\Pi}^{\gtrless}_{\boldsymbol{p} - \boldsymbol{p}' } (t,t') \cdot
              G^{\gtrless}_{\boldsymbol{p}'                  } (t,t'), \label{eq:sigma_2B_glsgt}
        \\
    \Pi^{\gtrless}_{\boldsymbol{q}}(t,t')
        & =
        \tfrac{1}{N} \sum_{\boldsymbol{p}''}
            G^{\gtrless}_{\boldsymbol{p}''               } (t,t') \cdot
            G^{\lessgtr}_{\boldsymbol{p}''-\boldsymbol{q}} (t',t),
\end{align}
where ``$\cdot$'' stands
for the multiplication of $2 \times 2$ matrices in spin space, and
$\bar{X}_{\boldsymbol{p}}(t,t') = \operatorname{tr} [X_{\boldsymbol{p}}(t,t')] - X_{\boldsymbol{p}}(t,t')$.

Using Eq.~(\ref{eq:sigma_2B_glsgt}), a modified form of Eq.~(\ref{eq:GR_from_GS}) with $G$ replaced by $\Sigma$, and the definitions below Eq.~(\ref{eq:RAK_matrix_def}), we compute the second-order contribution to the self-energy $\hat{\Sigma}^{(2)}$.
Additionally, we include the first-order (Hartree-Fock) self-energy contribution (retarded and advanced):
\begin{gather}
  \Sigma^{(\mathrm{HF})}_{\boldsymbol{p}} (t,t')
        = -i \tfrac{U}{N} \delta(t-t') \sum_{\boldsymbol{p}'} \bar{G}^{<}_{\boldsymbol{p}'} (t,t).
\end{gather}
We summarize the self-energy approximation used in this work as: $\hat{\Sigma} = \hat{\Sigma}^{(2)}+\left(\begin{smallmatrix}\Sigma^{(\mathrm{HF})}&0\\0&\Sigma^{(\mathrm{HF})}\end{smallmatrix}\right)$.

\begin{figure}[t]
        \includegraphics[width=\columnwidth]{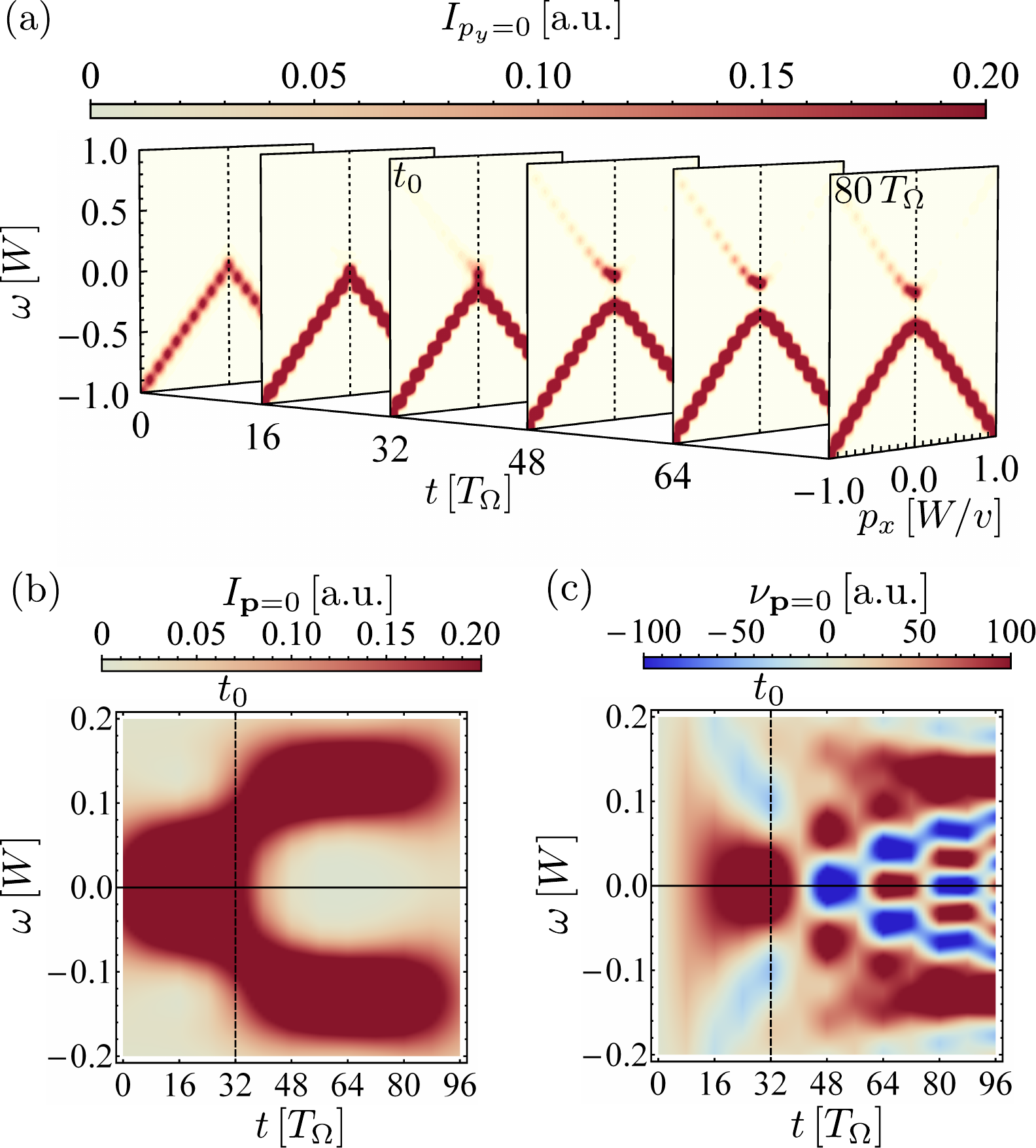}
\caption{
Dynamics of Floquet gap opening.
For $t<t_{0}$ the driving is off and thus the spectrum is gapless.
a) TRARPES signal, Eq.~(\ref{eq:trARPES_def}), at $p_y=0$, for a probe pulse width $\sigma = 16 T_{\Omega}$. Due to the initial band degeneracy a small density of particles is excited to the upper band after the quench.
b) TRARPES signal at $\boldsymbol{p}=0$.
The extracted value of the Floquet gap at the end of the run is $\Delta=0.115W$, which is enhanced compared to its value in the noninteracting case, $\Delta_0=0.088W$.
c) Generalized spectral function, Eq.~(\ref{eq:gen_Aspec_def}), at $\boldsymbol{p}=0$. $\operatorname{Sinc}$-like peaks at the emergent quasienergy band edges yield subdominant oscillations inside the gap.
Parameter values for all panels: $\Omega = 4W$, $A=0.6$, $U=0.4W$.
}
\label{fig:gap_opening}
\end{figure}

{\it Results.---}
We numerically solve Eq.~(\ref{eq:dyson_RAK}) for the dynamics, within the 2BA.
For the initial state before the driving quench, we wish to prepare the ground (or a low-temperature) state of the interacting, non-driven system.
To this end, we start with the {\it non-interacting} ground state at half-filling, and evolve with the full, static ($\mathcal{A} = 0$), interacting Hamiltonian for an equilibration time $t_0$~\footnote{Through this initialization procedure, the system thermalizes to a low (but nonzero) temperature equilibrium state.}.
The drive is then switched on for a time $t_{\mathrm{dr}}$.
For the subsequently discussed simulations, we used a grid with $21 \times 21$ points in momentum space, an equilibration time of $t_{0} = 32 T_{\Omega}$, and a run-time of $t_{\mathrm{dr}} = 64 T_{\Omega}$, where $T_{\Omega} = \frac{2\pi}{\Omega}$ is the period of the drive.

We first focus on the dynamics of the Floquet gap opening.
In Fig.~\ref{fig:gap_opening}a we show snapshots of the time evolution of the TRARPES signal at $p_y=0$, $I_{p_y=0}(\omega,t)$, see Eq.~(\ref{eq:trARPES_def}).
As the gap opens, the upper band is occupied mostly around $\boldsymbol{p}=0$ due to the initial degeneracy.
In Fig.~\ref{fig:gap_opening}b we illustrate the gap opening by focusing on the time evolution of the same signal at $\boldsymbol{p}=0$.
In the equilibration step, $t<t_0$, the gap remains closed and
$I_{\boldsymbol{p}=0}(\omega,t)$ is peaked at $\omega=0$.
At $t=t_0$, when the drive is switched on, the gap begins to emerge, saturating at a magnitude that we define as the many-body Floquet gap, $\Delta$.
The observed time scale for this gap opening is limited by the temporal width of the Gaussian probe signal, $\sigma$.

In Fig.~\ref{fig:gap_opening}c we show the same process in terms of the generalized spectral function, $\nu_{\boldsymbol{p}=0}(\omega,t)$, see Eq.~(\ref{eq:gen_Aspec_def}).
While the spectral function of an equilibrium system is always positive, its generalization to the non-equilibrium setting may be both positive and negative.
Due to the sudden quench of the drive, the observed peaks of $\nu_{\boldsymbol{p}=0}(\omega,t)$ corresponding to the upper and lower bands are sinc-like functions; their overlap gives rise to the checkerboard pattern seen around zero frequency for times after the driving is turned on.
The period of oscillations on the $\omega = 0$ axis for $t > t_0$ is set by the induced gap. We extract gap values from this oscillation period and by fitting sinc functions to the full signal at late times.  These values are consistent with that extracted from the splitting between peaks in the TRARPES signal.

\begin{figure}[t]
        \includegraphics[width=\columnwidth]{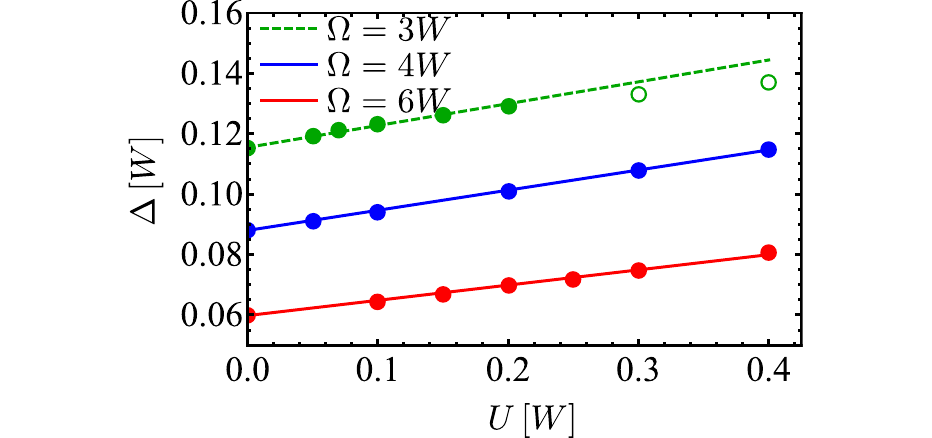}
\caption{
Many-body Floquet gap $\Delta$ (extracted from the TRARPES signal at the end of each run) as a function of interaction strength $U$ for $\Omega=3W,4W,6W$ and $A=0.6$. The gap renormalization is approximately linear in the interaction strength, reaching up to $35\%$ in the chosen parameter range. The values of $\Delta$ at high heating rates cannot be extracted unambiguously (empty circles). The fitted lines correspond to functions $\Delta = \Delta_0(1 + C U/W)$ where $\Delta_0$ are the respective non-interacting gap values and $C=0.63(\Omega=3W),0.76(\Omega=4W),0.84(\Omega=6W)$.
Note that the slope $C$ slightly increases for increasing driving frequency $\Omega$, i.e., for decreasing $\Delta_0$.
}
\label{fig:gap_vs_interaction}
\end{figure}

When interactions are strong, the magnitude of the Floquet gap $\Delta$ may be considerably enhanced as compared to the non-interacting case, see Fig.~\ref{fig:gap_vs_interaction}.
The increase of $\Delta$ has an approximately linear dependence on the interaction strength.
The slope in this linear dependence, normalized by the non-interacting Floquet gap $\Delta_0$, is a slowly increasing function of $\Omega$, and hence it is a slowly decreasing function of $\Delta_0$.
This behavior is consistent with the analytical estimate $\Delta/\Delta_0 \approx 1+\tfrac{\alpha}{2}\ln(D/\Delta_0)$, derived for an equilibrium 2D Dirac system at low temperature, with bandwidth $D$, bare gap $\Delta_0$, and dimensionless interaction strength $\alpha$~\cite{Kotov2012}.
(Here one may expect the renormalization to be cut off by finite effective temperature of the pre-thermal state; our data do not provide sufficient resolution to characterize this relation further.)
Comparing with Hartree-Fock level simulations, we observe that correlations captured at second-Born level slightly reduce the gap as compared with its value within the Hartree-Fock approximation.

\begin{figure}[t]
        \includegraphics[width=\columnwidth]{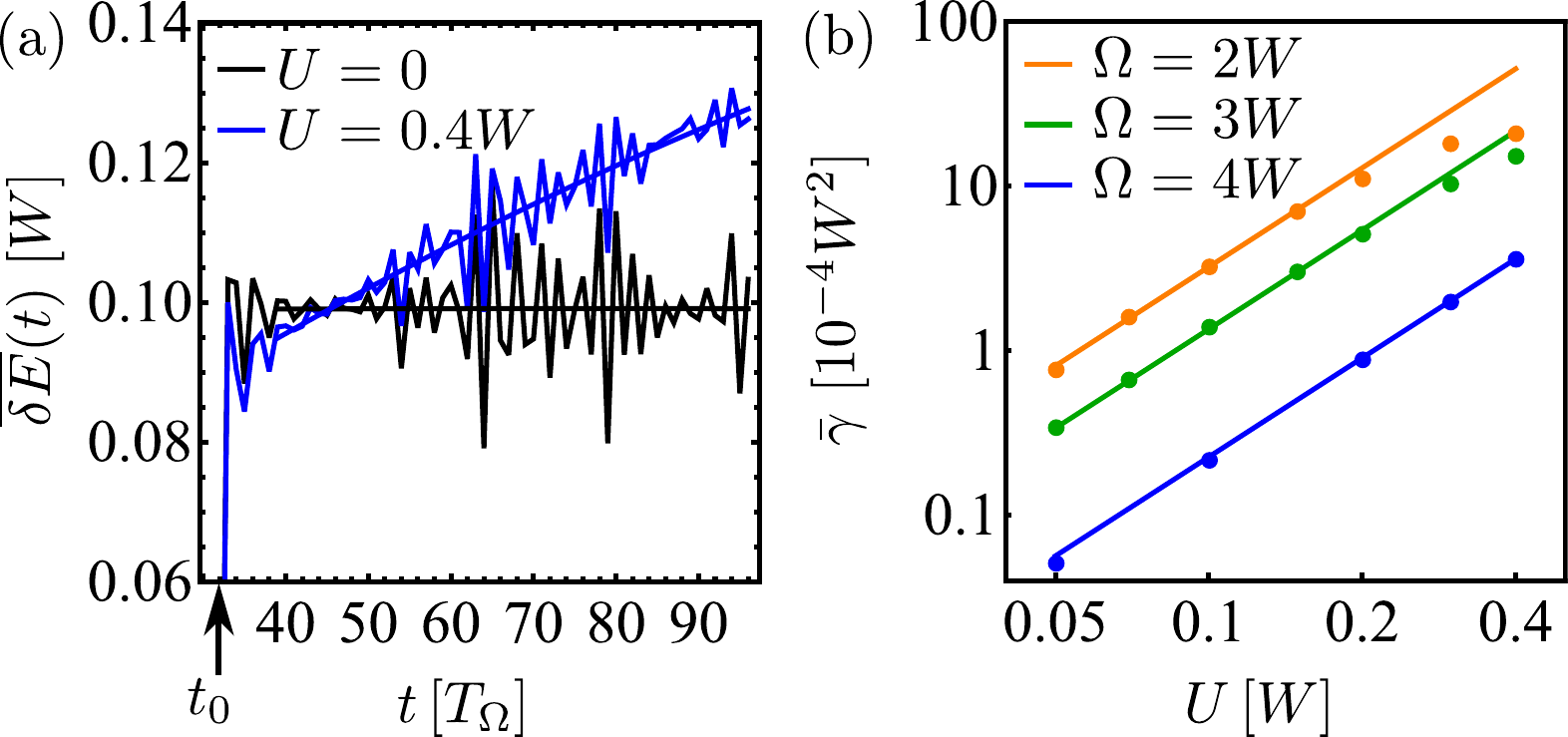}
\caption{ Analysis of heating in the high frequency regime, for $A=0.6$.
a) Period-averaged absorbed energy $\overline{\delta E}(t)$ as a function of time for $\Omega=4W$.
Curves for both the non-interacting ($U=0$, black) and interacting ($U=0.4W$, blue) cases oscillate, but only the latter has an overall slope which we extract as the effective energy absorption rate, $\bar{\gamma}$.
b) Effective energy absorption rate $\bar{\gamma}$ as a function of interaction (log-log plot) for $\Omega=2W,3W,4W$. At low heating rates we find $\bar{\gamma}\sim U^2$ (fitted lines). Note the quick decay of $\bar{\gamma}$ with increasing $\Omega$.
}
\label{fig:gamma_vs_interaction}
\end{figure}

We further investigate the energy absorbed in the system, $\delta E$, by integrating the absorption rate $\gamma$ over time.
In both the interacting and non-interacting cases, $\delta E$ exhibits large oscillations in time.
However, while $\delta E(t)$ saturates with time in the non-interacting case, it continues to grow with a net positive slope in the interacting case~\footnote{We additionally confirmed that heating is not captured within the Hartree-Fock approximation, as expected.}.
We extract the net average growth rate of $\delta E$ via a linear fit to the simulation data, and use its value $\bar{\gamma}$ as a measure of the heating rate, see Fig.~\ref{fig:gamma_vs_interaction}.
The value of $\bar{\gamma}$ rapidly decreases with increasing $\Omega$, in qualitative agreement with the expected exponential suppression of absorption in the high frequency regime~\cite{Abanin2015}.
Moreover, $\bar{\gamma}$ has a power-law dependence on U, $\bar{\gamma}\sim U^2$.

The value of $\bar{\gamma}$ can be translated into a heating time scale by estimating the time needed to absorb an energy of order $W$, $t_{\mathrm{heat}}=W/\bar{\gamma}$.
At frequencies of order a few times the band width,
we find that $t_{\mathrm{heat}}\sim (10^3-10^5)W^{-1}$ is well beyond the dynamical time scale of the gap opening. Thus there is an extended time window in which physical observables reach quasi-equilibrated values while the heating has not yet washed out the relevant low-energy properties of the system.

{\it Discussion.---}
This work is a first step in understanding the transient dynamics of gap opening in interacting, periodically driven systems.
The simple model that we employ allows us to explore the competition between the various processes at play in such systems.
We now briefly discuss how this model relates to more realistic systems that may be studied in experiments.

Our model includes a single species of 2D Dirac fermions.
Qualitatively, we do not expect the presence of additional species to significantly affect our results.

In order to avoid rapid heating under strong driving and interactions, the driving should not be resonant (i.e., $\Omega$ should remain larger than the bandwidth).
Since the Floquet gap decreases with $1/\Omega$, going to high frequencies also necessitates large amplitudes.
Thus small bandwidth Dirac systems, such as those on the surfaces of 3D topological insulators (TIs), appear to be most favorable.

In a small band gap TI, the driving frequency may easily be larger than the surface state bandwidth, and the driving amplitude may also be significant on this scale (as in our simulations).
However, in this situation the driving frequency will be greater than the bulk band gap, and its effect on the bulk spectrum and excitations must be taken into account.
Studying these effects and investigating the role of interactions in pump-probe experiments in specific materials are important directions for further work.

{\it Acknowledgments.---}
We thank D. Abanin, N. Lindner, E. Berg, G. Refael, A. Rosch and M. Genske for fruitful discussions.
We gratefully acknowledge the support of the European Research Council (ERC) under the European Union Horizon 2020 Research and Innovation Programme (Grant Agreement No. 678862), and the Villum Foundation.
%

\end{document}